\documentclass[fleqn,10pt]{wlscirep}
\usepackage[utf8]{inputenc}
\usepackage[T1]{fontenc}
\usepackage{bm}
\usepackage{comment}
\usepackage{braket}

\title{Quantum Annealing Enhanced Markov-Chain Monte Carlo}

\author[1]{Shunta Arai}
\author[2,3]{Tadashi Kadowaki}
\affil[1]{Institute of Science Tokyo, Ookayama, Tokyo 152-8550, Japan}
\affil[2]{Global R\&D Center for Business by Quantum-AI Technology, National Institute of Advanced
Industrial Science and Technology, Ibaraki, Japan}

\affil[3]{DENSO CORPORATION, 1-1-4, Haneda Airport, Ota-ku, Tokyo 144-0041, Japan}

\begin{abstract}
In this study, we propose quantum annealing-enhanced Markov Chain Monte Carlo (QAEMCMC), where QA is integrated into the MCMC subroutine. 
QA efficiently explores low-energy configurations and overcomes local minima, enabling the generation of proposal states with a high acceptance probability. 
We benchmark QAEMCMC for the Sherrington-Kirkpatrick model and demonstrate its superior performance over the classical MCMC method. Our results reveal larger spectral gaps, faster convergence of energy observables, and reduced total variation distance between the empirical and target distributions.
QAEMCMC accelerates MCMC and provides an efficient method for complex systems, paving the way for scalable quantum-assisted sampling strategies.
\end{abstract}
\begin{document}

\flushbottom
\maketitle

\thispagestyle{empty}

\noindent
\section*{Introduction}
Sampling from a Gibbs-Boltzmann distribution has numerous practical applications across various fields, such as combinatorial optimization \cite{Kirkpatrick_1983}, biology \cite{Li_1989}, and machine learning \cite{Andrieu_2003}.
 A widely used approach is the Markov-Chain Monte Carlo (MCMC) method \cite{levin2017markov} with the Metropolis-Hastings (M-H) acceptance probability \cite{Metropolis_1053}.
In this scheme, we generate a proposal state from any probability distribution and decide whether the proposal state is accepted or not, based on the M-H acceptance probability. 
This process is repeated until the samples can be considered to come from a stationary distribution. 
Many iterations are needed to reach the stationary distribution for complex energy landscapes like spin glass \cite{Binder_1986}. 
Therefore, faster convergence of MCMC is a crucial problem.

To achieve faster convergence of MCMC, a well-known approach is the cluster update \cite{Swendsen_1987, Wolff_1989}, which involves updating multiple spins at once and is more efficient than local update.
Cluster update is a technique primarily developed in pure systems. However, they are ineffective in spin glass due to frustration, which yields many local minima in the energy landscape \cite{Houdayer_2001}. 
They cannot overcome the slow relaxation in low temperature. 
For this reason, the extended ensemble methods such as exchange Monte Carlo \cite{Hukushima_1996} and parallel tempering \cite{Earl_2005} and the efficient sequential Monte Carlo algorithm like population annealing \cite{hukushima_2003} have been proposed to evade the slow relaxation for spin glass.

Acceleration of the dynamics of MCMC for spin glass is a challenging task. In a previous study, machine learning has been applied to accelerate the dynamics of MCMC.
One of the pioneering research is the self-learning Monte Carlo (SLMC) method \cite{Liu_2017a, Huang_2017}.
In SLMC, the effective model trained from training data has been introduced to accelerate the dynamics of MCMC.
The proposal from the effective model has yielded the global update and brought about faster convergence of MCMC.
This pioneering research has led to the development of \textit{Neural MCMC} (NMCMC) algorithm \cite{Wu2018,McNaughton_2020,nicoli_2020,Wu_2021,Bia_2021}.
In NMCMC, an autoregressive neural network, which has a high representation power and whose probability can be computed easily without evaluating the partition function, has often been applied as the effective model. 
Although NMCMC could provide a computational speed-up over the MCMC with a local update for the ferromagnetic Ising model and two-dimensional (2d)-Edward Anderson model, NMCMC has not worked in the
antiferromagnetic Potts model on a random graph, which is the representative example of the hard problem with random first-order phase transition \cite{Ciarella_2023}.

To accelerate the dynamics of MCMC, an alternative approach has been proposed called the \textit{quantum-enhanced MCMC} (QEMCMC) \cite{Layden_2023}.
In QEMCMC,  gate-based quantum computers generate new proposal states using quantum time evolution.
The classical computer subsequently determines whether the proposal state is accepted or not.
The previous study numerically has demonstrated that 
this hybrid quantum and classical algorithm converged faster than the classical MCMC method with uniform and local updates for the small spin systems \cite{Layden_2023}.
The proposal states had energy similar to the current state with a larger Hamming distance than the local update. 
This characteristic of samples obtained from QEMCMC led to faster convergence of MCMC. 
Moreover, QEMCMC has been extended to a coarse-grained approach for larger size problems than the size of the quantum computer \cite{stuart_2024}, variational methods for adjusting hyperparameters \cite{Nakano_2024}, and continuous systems \cite{Owen_2024}.
In the recent theoretical analysis, QEMCMC has not yielded a quantum speed-up over classical sampling on the masked item sampling problem \cite{Orfi_pra_2024a}. A bottleneck analysis has indicated the region where quantum improvement may exist \cite{Orfi_pra_2024b}. 

In addition to the gate-based quantum computer, quantum time evolution can be realized in a quantum annealer \cite{Dwave2010a, Dwave2010b, Dwave2010c, Dickson_2013, Lanting_2014}.
The quantum annealer is the physical implementation of quantum annealing (QA), which is the metaheuristic for combinatorial optimization problems \cite{Kadowaki_1998, Farhi_2001, Das_2008, Albash_2018,  Hauke_2020, Crosson_2021}.
The primary usage of the quantum annealer is divided into \textit{optimization} and \textit{sampling}.
For the optimization task, the quantum annealer has been applied to various combinatorial optimization problems  \cite{oshiyama_2022, Yarkoni_2022}.
For the sampling task, the quantum annealer is used to generate samples from the approximated Gibbs-Boltzmann distribution governed at a hardware-specific effective temperature \cite{Amin_2015, Vuffray_2022, Nelson_2022}. 
Recently, the quantum annealer has been
applied to various quantum simulations at a finite temperature such as three-dimensional spin glass \cite{Harris_2018}, fully frustrated square-octagonal lattice \cite{andrew_2018}, and spin ice \cite{andrew_2021a}.
Hardware development of the quantum annealer enables the operation in a coherent regime \cite{king_2022,king_2023}.

The examples presented above utilize the outputs from the quantum annealer directly.
In this paper, we propose the usage of QA in the MCMC subroutine. 
We focus on the features that QA can yield low-energy configurations and escape local minima efficiently.
Inspired by QEMCMC \cite{Layden_2023},  we generate the proposal state by QA. 
The classical computer accepts or rejects the proposal state with the M-H acceptance probability.
Because the resultant Markov chain is satisfied with the detailed balance condition, the stationary distribution obtained by QAEMCMC matches the target distribution.
In the previous study \cite{Giuseppe_2023}, the quantum annealer has been utilized for a sampler of the training data for NMCMC.
The NMCMC sampler, which mimics the output distribution of the quantum annealer, has been used as the proposal distribution. 
In this study, we focus on the potential of the direct output distribution obtained by QA in the MCMC subroutine.
We investigate how the advantage of quantum fluctuations at low temperatures affects the sampling with the MCMC procedure.
Following the previous study \cite{Layden_2023}, we consider the Sherrington-Kirkpatrick (SK) model \cite{Sherrington_1975} as a benchmark problem.
We show the analysis of the absolute spectral gap and the behaviors of observables. 
We demonstrate that  QAEMCMC yields significant spectral gaps and faster convergence of the Markov chain than the classical updates.

This paper is organized as follows. 
In the next section, we explain the classical MCMC method and QAEMCMC. 
In the following section, we present the numerical simulations of QAEMCMC for the SK model.
The final section is dedicated to discussing and summarizing the results.
\section*{Methods}
\label{sec::sec2}
In this section, we explain the mathematical formulation of MCMC and the proposed method.

\subsection*{Markov-Chain Monte Carlo}
\label{sec::sec21}
MCMC starts from a trivial spin configuration $\bm{\sigma}$ and constructs a sequence of states, which is called a Markov chain, by updating the current state with the transition probability $P(\bm{\sigma}'|\bm{\sigma})$.
The transition probability is designed so that the stationary distribution corresponds to the target distribution $\mu(\bm{\sigma})$.
Then, the resultant Markov chain can be regarded as samples from the target distribution if MCMC iterates sufficiently. The sufficient condition to converge the Markov chain to the target distribution is to satisfy the detailed balance condition : 
\begin{equation}
P(\bm{\sigma}'|\bm{\sigma})\mu(\bm{\sigma})=P(\bm{\sigma}|\bm{\sigma}')\mu(\bm{\sigma}'),
\label{eq1}
\end{equation}
for all $\bm{\sigma}\neq\bm{\sigma}'$. 

The  popular approach to construct $P(\bm{\sigma}'|\bm{\sigma})$ is the M-H method \cite{Metropolis_1053}.
The M-H method decomposes $P(\bm{\sigma}'|\bm{\sigma})$ into the acceptance probability $A(\bm{\sigma}'|\bm{\sigma})$ and the proposed probability $Q(\bm{\sigma}'|\bm{\sigma})$ as $P(\bm{\sigma}'|\bm{\sigma})=A(\bm{\sigma}'|\bm{\sigma})Q(\bm{\sigma}'|\bm{\sigma})$.
At first, we sample the proposal state $\bm{\sigma}'$ from $Q(\bm{\sigma}'|\bm{\sigma})$. Next, we decide whether the current proposal is accepted or rejected with the M-H probability as 
\begin{equation}
   A(\bm{\sigma}'|\bm{\sigma})= \mathrm{min}\left(1,\frac{\mu(\bm{\sigma}')}{\mu(\bm{\sigma})}\frac{Q(\bm{\sigma}|\bm{\sigma}')}{Q(\bm{\sigma}'|\bm{\sigma})}\right).
   \label{eq2}
\end{equation}

For a binary system $\bm{\sigma}\in \{1,-1\}^N$ where $N$ is the system size, the local update is often applied. 
This involves choosing a random spin index $i\in[1, N]$ of the current spin configuration and flipping it as $\sigma_i' \leftarrow -\sigma_i $.
The mathematical representation of $Q(\bm{\sigma}'|\bm{\sigma})$ for the local update can be described as 
\begin{equation}
    Q_{\mathrm{local}}(\bm{\sigma}'|\bm{\sigma})=\begin{cases}\frac{1}{N} \quad &(d(\bm{\sigma}',\bm{\sigma})=1)\\
    0\quad \quad &(otherwise)
\end{cases}
    \label{eq3},
\end{equation}
where $d(\bm{\sigma}',\bm{\sigma})=\sum_{i=1}^N|\sigma_i'-\sigma_i|$ is the Hamming distance between the current and proposed states.
In the local update, Eq. \eqref{eq2} is easily calculated since the ratio of the Boltzmann factor $\mu(\bm{\sigma})/\mu(\bm{\sigma}')$ can be efficiently computed.
It is noted that the local update often fails to escape local minima, and the resultant acceptance probability is very low in low-temperature regions \cite{levin2017markov}.
Another common approach is the uniform update, which employs uniform sampling from
\begin{equation}
      Q_{\mathrm{uniform}}(\bm{\sigma})=
      \frac{1}{2^N}.
      \label{eq4}
\end{equation}
In the case that the proposed state is independent of the current state, the M-H acceptance probability can be extended as 
\begin{equation}
   A(\bm{\sigma}'|\bm{\sigma})= \mathrm{min}\left(1,\frac{\mu(\bm{\sigma}')}{\mu(\bm{\sigma})}\frac{Q(\bm{\sigma})}{Q(\bm{\sigma}')}\right),
   \label{eq5}
\end{equation}
which is called as \textit{Metoplized independent sampling} \cite{Liu_1996}.
It is noted that a simple calculation confirms the detailed balance condition with Eq. \eqref{eq5}.
While the uniform update may give rise to the proposal states with a large Hamming distance and lead to escape from the local minima, the acceptance probability becomes low. 
As a result, the small acceptance probability deteriorates the convergence rate of the Markov chain. 

Mixing time $\tau_\epsilon$ is utilized to quantify the convergence rate of the Markov chain.
It is defined as the minimum number of iterations needed for the probability distribution obtained by the Markov chain to get within $\epsilon >0$ of $\mu(\bm{\sigma})$ in total variation distance for any initial distribution as 
\begin{equation}
(1-\delta^{-1})\ln 2\epsilon \leq \tau_\epsilon \leq -\delta^{-1}\ln \left(\epsilon \mathrm{min}_{\bm{\sigma}}\mu(\bm{\sigma})\right)
\label{mixing_time_ineq}
\end{equation}
where $\delta$ is the absolute spectral gap as $\delta = 1- \mathrm{max}_{\lambda\neq 1}\lambda$ and $\lambda$ is the eigenvalues of $P(\bm{\sigma}'|\bm{\sigma})$ \cite{levin2017markov}.
Equation. \eqref{mixing_time_ineq} shows that the mixing time is bounded by the absolute spectral gap, and a larger spectral gap results in a shorter mixing time.
The absolute spectral gap depends on the choice of $Q(\bm{\sigma}'|\bm{\sigma})$ in the M-H method. 
Choosing a proposal probability that increases the absolute spectral gap leads to faster convergence of MCMC.

\subsection*{QA-enhanced MCMC}
\label{sec::sec22}
In QA, the Hamiltonian can be constructed as 
\begin{equation}
\hat{H}=\mathcal{A}(s(t))\hat{H}_0(\bm{\hat{\sigma}}^z)-\mathcal{B}(s(t))\sum_{i=1}^N\hat{\sigma}_i^x
\label{eq7}
\end{equation}
where $\hat{\sigma}_i^{z}$ and $\hat{\sigma}_i^{x}$ are the Pauli operators acting on the site $i$, and $\hat{H}_0(\bm{\hat{\sigma}}^z)$ is the target Hamiltonian whose eigenvalues correspond to the energy of the Gibbs-Boltzmann 
distribution 
\begin{equation}
  \mu(\bm{\sigma})=\frac{1}{Z}\bra{\bm{\sigma}} \exp(- \beta  \hat{H}_0(\bm{\hat{\sigma}}^z))\ket{\bm{\sigma}}.
  \label{eq8}
\end{equation}
The partition function is defined as 
$Z=\sum_{\bm{\sigma}}\bra{\bm{\sigma}}   \exp(- \beta  \hat{H}_0(\bm{\hat{\sigma}}^z))\ket{\bm{\sigma}}$ and $\beta=1/T$ is the inverse temperature.
The annealing schedule function $\mathcal{A}(s(t))$ and $\mathcal{B}(s(t))$ interpolate two Hamiltonians as a function of annealing parameter $s(t)$. We adopt the linear schedule as $\mathcal{A}(s(t))=s(t)$ and   $\mathcal{B}(s(t))=1-s(t)$.
In the vanilla QA, the initial state is the superposition state, which is the ground state of the second term in Eq. \eqref{eq7}, and evolves following the Schr\"odinger dynamics.
The annealing parameter $s(t)$ increases monotonically from $s(t=0)=0$ to  $s(t=\tau)=1$.
The annealing time $\tau$ determines the period of total time evolution.
If the dynamics evolves adiabatically, the final state concentrates on the ground state of $\hat{H}_0(\bm{\hat{\sigma}}^z)$
in the limit of $\tau \rightarrow \infty$.

\begin{figure}[t]
\centering
\includegraphics[width=170mm]{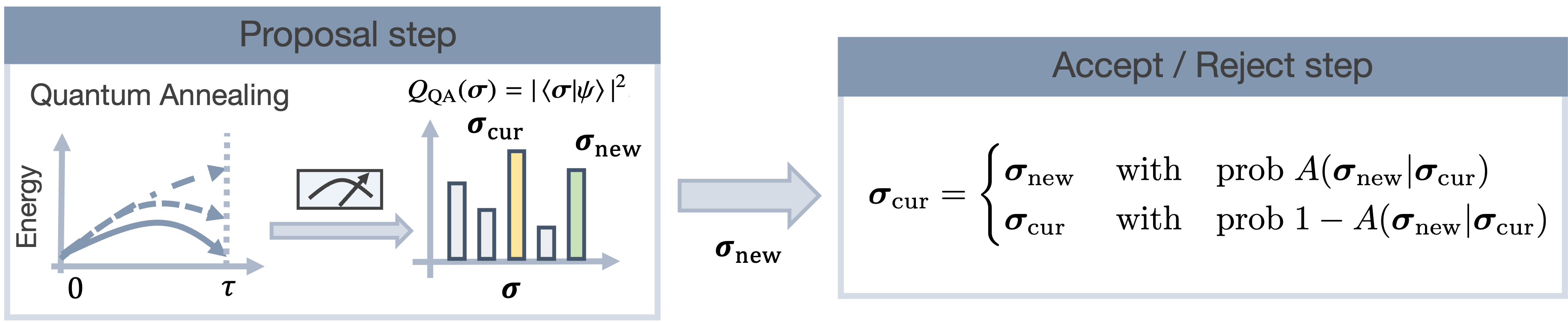}
\caption{The schematic diagram of the QAEMCMC protocol. 
}
\label{fig:fig1}
\end{figure}

In the short or intermediate annealing time, the final states deviate from the ground state, and various eigenstates can be found. 
The classical spin configuration samples from the final state of the wavefunction $\ket{\psi}$ with the probability
\begin{equation}
    Q_{\mathrm{QA}}(\bm{\sigma})=|\braket{\bm{\sigma}|\psi}|^2
    \label{eq9}.
\end{equation}
For the large but not infinite annealing time, the probability concentrates around the ground state of the target Hamiltonian, and we can find the low-energy configurations.
We utilize this diabatic effect of QA to generate the proposal state in the MCMC subroutine.
 According to the standard protocol of the M-H method, the proposed state is sampled from Eq. \eqref{eq9} and decided whether the proposed state is accepted or not.
This QA update and acceptance or rejection steps are iterated.
We refer to the above procedure as \textit{QA-ehnhanced MCMC} (QAEMCMC).
The schematic diagram of the QAEMCMC protocol is plotted in Fig. \ref{fig:fig1}.
\section*{Results}
\label{sec:sec3}
In this section, we show the results of the SK model as 
\begin{equation}
    \hat{H}_0(\hat{\bm{\sigma}}^z)=\sum_{i< j}J_{ij}\hat{\sigma}_i^z\hat{\sigma}_j^z+\sum_{i=1}^Nh_i\hat{\sigma}_i^z
\end{equation}
where $J_{ij}$ and $h_i$ are sampled from the Gaussian distribution with zero mean and unit variance.
The model is the same as used in the previous studies \cite{Layden_2023, Nakano_2024}.
\begin{figure}[t]
\centering
\includegraphics[width=175mm]{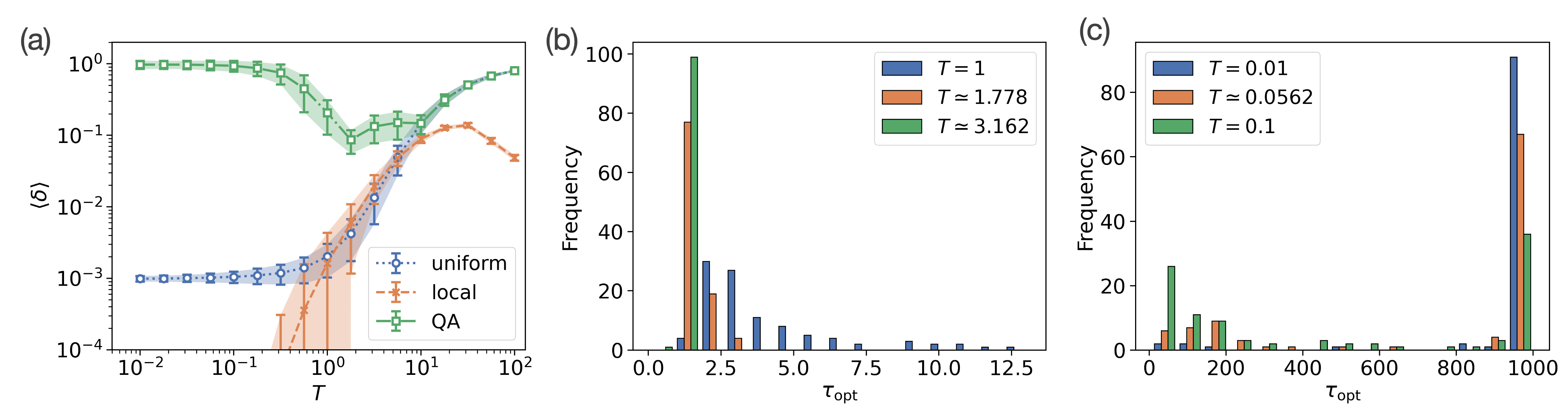}
\caption{(a): The absolute spectral gap as a function of temperature for the fixed $N=10$.
Each marker represents different update schemes.
The histogram of the optimized annealing time for (b): intermediate temperature and (c): low temperature regions.
}
\label{fig:fig2}
\end{figure}

\begin{figure}[t]
\centering
\includegraphics[width=80mm]{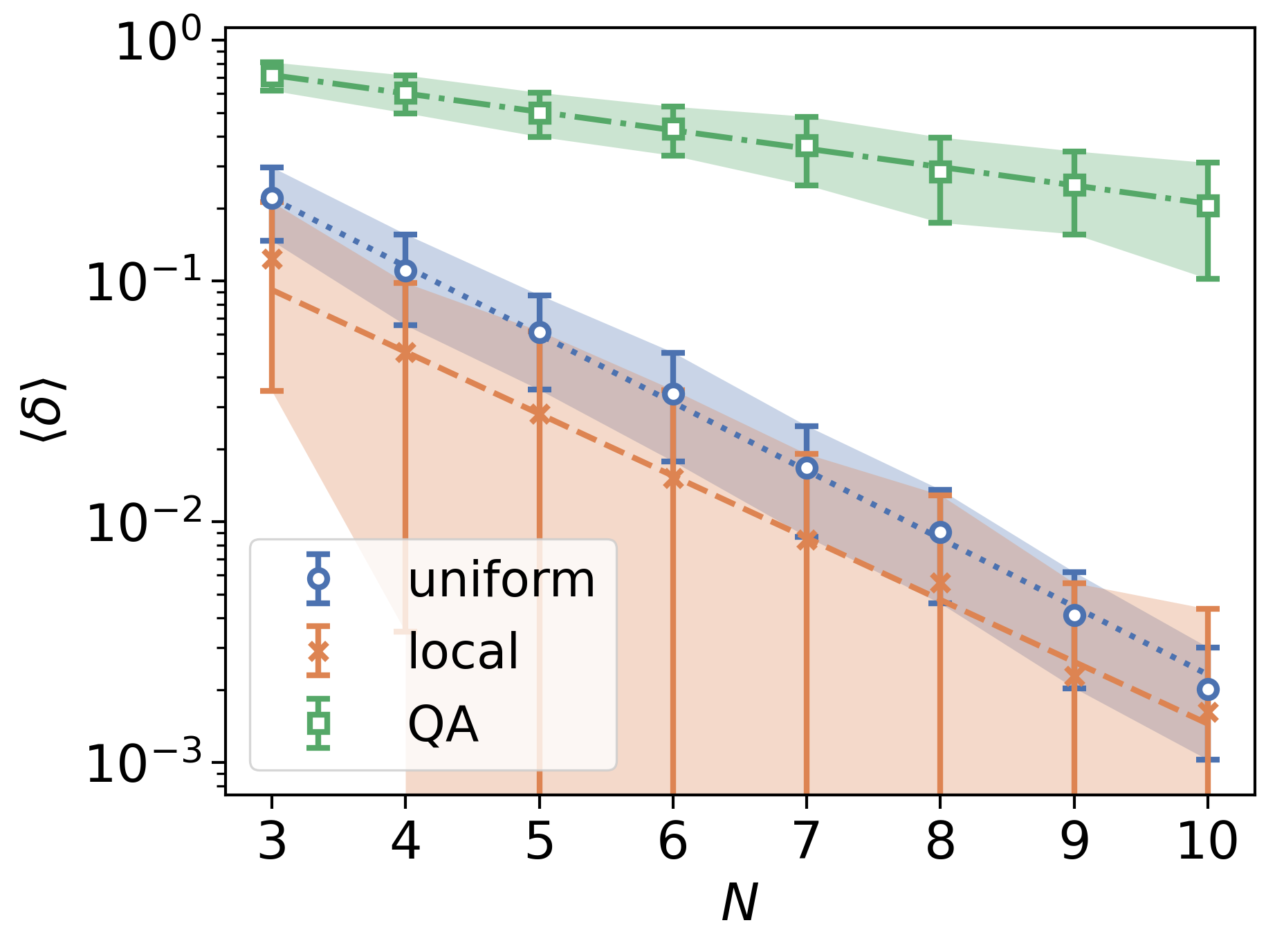}
\caption{
The system size dependence of the absolute spectral gap for the fixed $T=1$ with different update schemes.
}
\label{fig:fig3}
\end{figure}

We exhibit the absolute spectral gap $\langle \delta \rangle$ for $N=10$ as a function of temperature $T$ in Fig. \ref{fig:fig2} (a).
The bracket $\langle \cdot\rangle$ represents the average for the $100$ random instances.
The error bars and shaded regions represent the standard deviation. 
For each instance and $T$, 
we compute the eigenvalues of the transition probability. 
The QA simulation is carried out with the QuTip
 \cite{Johansso_2012a, Johansso_2012b}.
In the QA update, $\tau$ is tuned to maximize the absolute spectral gap by Optuna \cite{Akiba_2019} for $T<10$.
In high-temperature region $T\geq 10$, we fix $\tau=0.01$ with the prior knowledge of the target distribution. 
In this region, the Gibbs-Boltzmann distribution approaches the uniform distribution.
The proposal distribution of the QA update with small $\tau$ approximately yields the uniform distribution since few transitions of the states from the initial superposition states occur.
As a result, the uniform and QA updates take the largest $\langle \delta \rangle$ because the proposal distribution is similar to the target distribution.
For the local update, $\langle \delta \rangle$ shows the parabolic behavior. 
This behavior is reflected in the structure of eigenvalues of the transition probability and was also discussed in the previous study \cite{Layden_2023}.
In the intermediate region $1\leq T <10$, the QA update takes the larger $\langle \delta \rangle$ than those from the uniform and local updates. 
Figure \ref{fig:fig2} (b) exhibits the histogram of the optimized $\tau$ with $N=10$ in the intermediate temperature.
The broader histogram shows the strong dependence of the optimized $\tau$ on each instance.
In this region, the target distribution is a multi-modal distribution dominated by several low-energy states.
That leads to the difficulty in optimization of $\tau$.
As we increase the temperature, the histogram is concentrated on the small $\tau$.
This observation is consistent with the results that the target distribution can be approximated by short annealing times and yields a large spectral gap in the high-temperature region.
At low $T<1$, the uniform and local updates do not work due to the existence of the stable local minima and lead to a small $\langle \delta \rangle$.
This difficulty is alleviated by the QA update, enabling us to escape from the local minima.
Since $Q_{\mathrm{QA}}(\bm{\sigma})$ approximates the target distribution well,
$\langle \delta \rangle\simeq 1.0$ can be obtained.
Figure \ref{fig:fig2} (c) shows the histogram of the optimized $\tau$ with $N=10$ in the low-temperature region.
The lower the temperature is, the greater frequency is concentrated around the large $\tau=10^3$. 
The Gibbs-Boltzmann distribution is dominated by the ground state in the low-temperature region.
The original usage of QA for obtaining the ground state matches the situation in this low temperature.

\begin{table}[t]
  \centering
  \caption{The fitting exponents of the results in Fig. \ref{fig:fig3}.}
  \scalebox{1.3}{ 
  \begin{tabular}{|c||c|c|c|c|}  \hline
     update&$\langle \delta \rangle \sim 2^{-\alpha N}$
    \\ \hline 
    uniform &$\alpha=0.939\pm0.017$
    \\ \hline
    local &$\alpha=0.855\pm0.013$
    \\ \hline
    QA &$\alpha=0.254\pm0.005$
    \\ \hline
  \end{tabular}
    \label{table_1}
    }
\end{table}
Figure \ref{fig:fig3}  shows the dependence of $\langle \delta \rangle$ on $N$ for the fixed $T=1$. 
Compared to $\langle \delta \rangle$ obtained from the uniform and local updates, $\delta$ from the QA update decreases slowly. 
We fit $\langle \delta \rangle$ as $\langle \delta \rangle\sim 2^{-\alpha N}$ and exhibit the fitting exponents in Table \ref{table_1}. 
More than doubled quantum enhancement exists for the exponents. 
From the results of $\langle \delta \rangle$, mixing time obtained by the QA update becomes smaller than those obtained by the other updates.
The QA update reduces the effect of slow relaxation at a low temperature and accelerates the MCMC convergence.

\begin{figure*}[t]
\centering
\includegraphics[width=150mm]{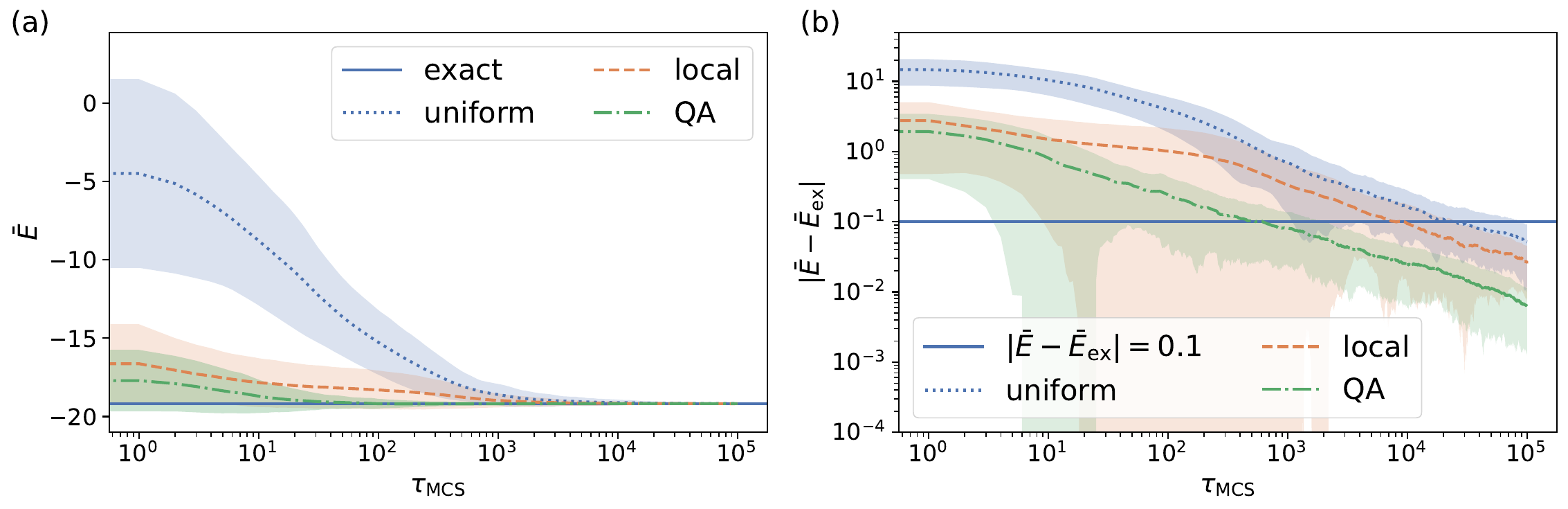}
\caption{(a): The dynamics of the averaged energy obtained at $T=1$ from MCMC. 
The horizontal axes represent the number of MCS.
The horizontal line denotes the exact equilibrium energy.
(b): The absolute error between the exact equilibrium energy and the empirical averages. 
The horizontal line represents $\Delta E =0.1$.
}
\label{fig:fig4}
\end{figure*}

To investigate the acceleration of the MCMC convergence, we demonstrate the dynamical behaviors of the averaged energy 
as a function of the Monte Carlo step (MCS) $\tau_{\mathrm{MCS}}$ represented as  
$\bar{E}=\sum_{i=1}^{\tau_{\mathrm{MCS}}}\bra{\bm{\sigma}^{(i)}}   \hat{H}_0(\bm{\hat{\sigma}}^z)\ket{\bm{\sigma}^{(i)}}/{\tau_{\mathrm{MCS}}}$ where $\bm{\sigma}^{(i)}$ is the spin configuration generated by each Monte Carlo step in Fig. \ref{fig:fig4} (a).
The experimental settings are as follows:
$N=10$ and $T=1$.
To demonstrate the performance of the QA update, we select a challenging problem instance used in Fig. \ref{fig:fig2}.
Its target distribution is multi-modal and dominated by five low energy states with $\mu(\bm{\sigma}) > 0.1$, which is difficult to sample efficiently.
Each line represents the averaged values computed from the independent $100$ MCMC runs. 
The shaded region represents the standard deviation.
The result obtained by the QA update converges the exact equilibrium energy $\bar{E}_{\mathrm{ex}}=\sum_{\bm{\sigma}}\mu(\bm{\sigma}) \bra{\bm{\sigma}}   \hat{H}_0(\bm{\hat{\sigma}}^z)\ket{\bm{\sigma}}$ faster than those by other updates.
Figure \ref{fig:fig4} (b) shows the absolute error between the averaged energy and the exact equilibrium energy as $ | \ \bar{E} - \bar{E}_{\mathrm{ex}}|$.
The horizontal line denotes $| \ \bar{E} - \bar{E}_{\mathrm{ex}}|=0.1$. 
The QA update achieves $| \ \bar{E} - \bar{E}_{\mathrm{ex}}|=0.1$ at $\tau_{\mathrm{MCS}}\simeq 618$.
The uniform and local updates reach at $\tau_{\mathrm{MCS}}\simeq 21232$ and $\tau_{\mathrm{MCS}}\simeq 7920$ respectively.
The QA update gives a better estimator, at least more than about twelve times faster than the classical updates.
Therefore, the QA update can realize efficient sampling of the low-energy states dominated by the Gibbs-Boltzmann distribution.

\begin{figure*}[t]
\centering
\includegraphics[width=150mm]{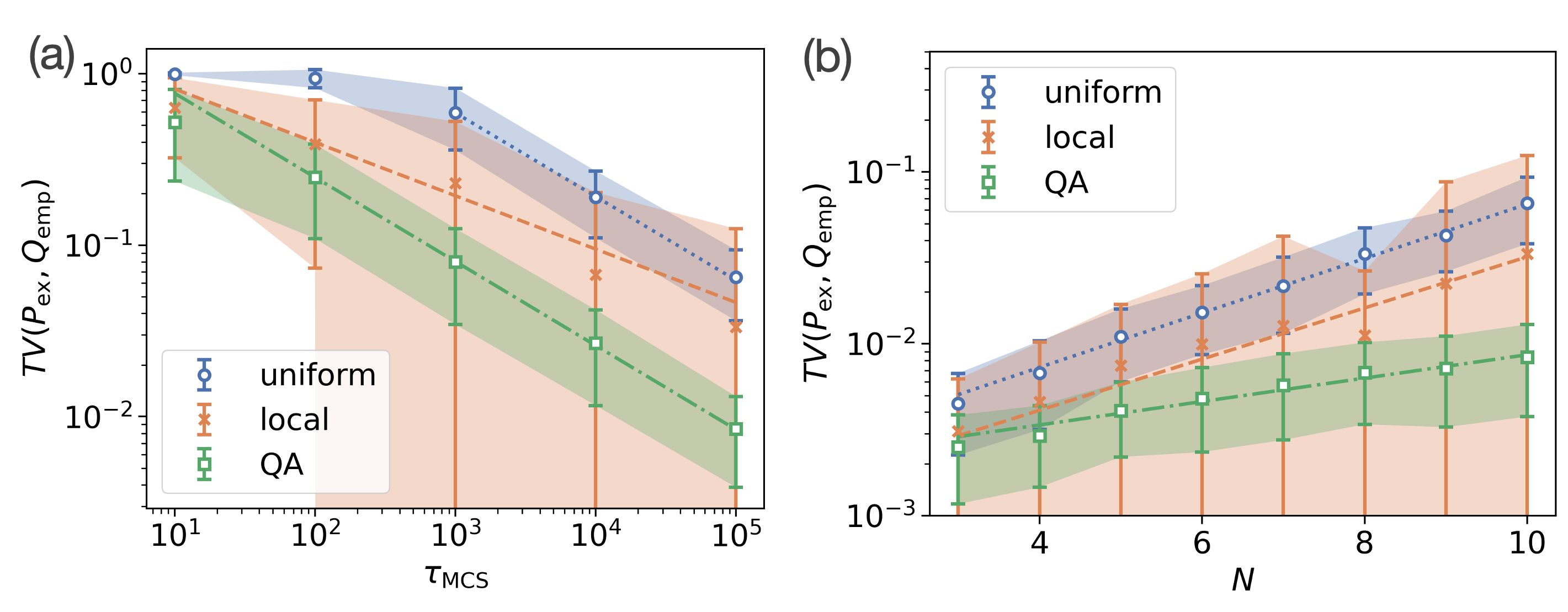}
\caption{(a): The Monte Carlo step dependence of the total variation distance between the Gibbs-Boltzmann distribution and the empirical probability distribution with fixed $T=1$ and $N=10$.
(b): the system size dependence of the total variation distance with fixed $T=1$ and $\tau_{\mathrm{MCS}}=10^5$.
}
\label{fig:fig5}
\end{figure*}

In addition to the efficacy of the sampling of the low-energy states, whether the empirical distribution $Q_{\mathrm{emp}}(\bm{\sigma})$ obtained by MCMC converges to the target distribution $P_{\mathrm{ex}}$ is also important in MCMC.
We compute the total variation distance 
between $P_{\mathrm{ex}}(\bm{\sigma})$ and  $Q_{\mathrm{emp}}(\bm{\sigma})$ as 
\begin{equation}
TV(P_{\mathrm{ex}},Q_{\mathrm{emp}})=\frac{1}{2}\sum_{\bm{\sigma}}|P_{\mathrm{ex}}(\bm{\sigma})-Q_{\mathrm{emp}}(\bm{\sigma})|_1.
\label{eq6}
\end{equation}
The experimental settings are the same as those in Fig. \ref{fig:fig4}. 
Figure \ref{fig:fig5} (a) demonstrates the dependence of $TV(P_{\mathrm{ex}},Q_{\mathrm{emp}})$ 
 on $\tau_{\mathrm{MCS}}$. 
 The scaling of $TV(P_{\mathrm{ex}},Q_{\mathrm{emp}})$ follows the power law decay as a function of $\tau_{\mathrm{MCS}}$ shown in Table \ref{table_2}.
 The QA update gives larger exponents than the local updates.
Even though the difference in the exponents between the QA update and the uniform update for $\tau_{\mathrm{MCS}}$ is small, the constant gap exists. 
Therefore, the QA update converges to the target distribution faster than other updates.
Figure \ref{fig:fig5} (b) exhibits the dependence of $TV(P_{\mathrm{ex}},Q_{\mathrm{emp}})$ on $N$ with $\tau_{\mathrm{MCS}}=10^5$.
The scaling exponents fitted by $2^{\gamma N}$ are shown in Table \ref{table_2}. 
The QA update yields smaller exponents than the classical updates.
The difference between the QA and other updates increases as we increase $N$.
For the large $N$, the QA update is effective. 
The acceleration of the MCMC convergence by the QA update can be seen from the dynamical simulations.

\begin{table}[t]
  \centering
  \caption{The scaling exponents of the total variation distance for the data in Fig. \ref{fig:fig5}.}
  \scalebox{1.3}{ 
  \begin{tabular}{|c||c|c|c|}  \hline
    update &$TV(P_{\mathrm{ex}},Q_{\mathrm{emp}}) \sim \tau_{\mathrm{MCS}}^{-\alpha}$ &$TV(P_{\mathrm{ex}},Q_{\mathrm{emp}}) \sim 2^{\gamma N }$
    \\ \hline 
    uniform &$\alpha=0.939\pm0.017$
    &$\gamma=0.526\pm0.017$
    \\ \hline
    local &$\alpha=0.855\pm0.013$
        &$\gamma=0.494\pm0.057$
    \\ \hline
    QA &$\alpha=0.254\pm0.005$
        &$\gamma=0.226\pm0.017$
    \\ \hline
  \end{tabular}
    \label{table_2}
    }
\end{table}

\begin{figure*}[t]
\centering
\includegraphics[width=175mm]{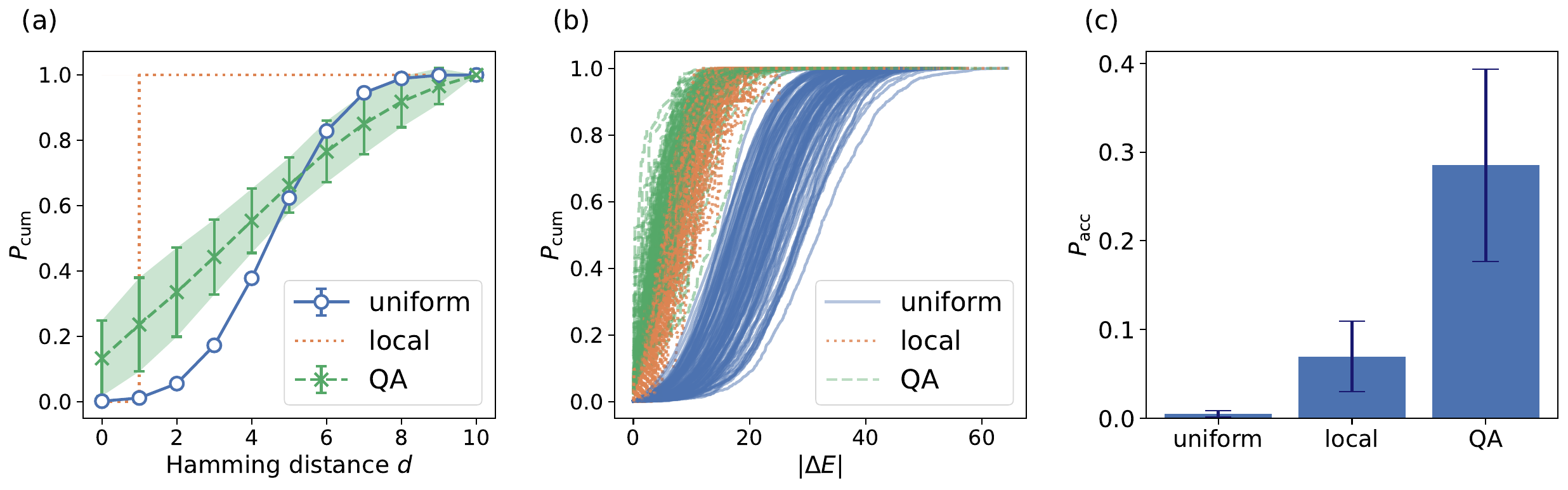}
\caption{The dependence of the cumulative probability distribution on (a): the Hamming distance and (b): the difference of the absolute energy between the current configurations and the proposed ones for the fixed $N=10$ and $T=1$. Each symbols represent the different update schemes.
(c): the acceptance probability for each update. The error bar represents the standard deviation.
}
\label{fig:fig6}
\end{figure*}

Finally, we investigate the features of samples obtained from different updates with the cumulative probability of the observables $\mathcal{O}(\bm{\sigma},\bm{\sigma}')$ .
For each Markov chain, we compute $\mathcal{O}(\bm{\sigma},\bm{\sigma}')$ with
 the current spin configuration $\bm{\sigma}$ and proposed one $\bm{\sigma}'$.
We focus on the cumulative probability of Hamming distance between two spin configurations and the absolute energy difference $|\Delta E|(\bm{\sigma},\bm{\sigma}')=|\bra{\bm{\sigma}}\hat{H}_0\ket{\bm{\sigma}}-\bra{\bm{\sigma}'}\hat{H}_0\ket{\bm{\sigma}'}|$.
The cumulative probability is calculated from $n_{\mathrm{all}}=10^5$ spin configurations as 
\begin{align}
   P_{\mathrm{cum}}(d)&=\sum_{x=0}^{d} P_{\mathrm{ham}}(x),\hspace{1mm}  P_{\mathrm{ham}}(x)=\frac{\# d(\bm{\sigma},\bm{\sigma}')=x}{n_{\mathrm{all}}},\\
    P_{\mathrm{cum}}(|\Delta E|)&=\int_{0}^{|\Delta E|} P_{|\Delta E|}(x)dx,\hspace{1mm}  P_{|\Delta E|}(x)=\frac{\# |\Delta E|(\bm{\sigma},\bm{\sigma}')=x}{n_{\mathrm{all}}},
\end{align}
where $\# \mathcal{O}(\bm{\sigma},\bm{\sigma}')=x$ denotes the number of samples satisfied with $\mathcal{O}(\bm{\sigma},\bm{\sigma}')=x$.
For the uniform and QA updates, we utilize the data with $\tau_{\mathrm{MCS}}=10^5$ in Fig. \ref{fig:fig5}.
Figure \ref{fig:fig6} (a) shows 
the dependence of $P_{\mathrm{cum}}(d)$ on Hamming distance.
Each marker denotes the average over $100$ different instances used in Figs. \ref{fig:fig2} and \ref{fig:fig5}.
The error bar and the shaded region denote the standard deviation.
As defined, the local update proposes the next state with $d=1$, while the proposal distribution of the uniform update concentrates around $d=5$, which is the highest number of states. 
This result is consistent with the result presented in the previous study \cite{Layden_2023}.
The QA update gives the samples with a smaller Hamming distance than the uniform update.
The main feature of the QA update is that we can obtain the samples with a large Hamming distance, which is far from the current state. 
Figure \ref{fig:fig6} (b) exhibits the dependence on the absolute difference of the energy between the current and proposed states represented as $|\Delta E|$. Each curve denotes the results for different instances.
In many instances, the QA update gives samples with a smaller  $|\Delta E|$ than the uniform and local updates. 
As shown in Fig. \ref{fig:fig6} (c), the small $|\Delta E|$ leads to a high acceptance probability. 
Therefore, the QA update can yield the samples with large $d$ and small $|\Delta E|$. 
We can interpret the QA update, which enables us to transition to the different local minima.
\section*{Discussion and Conclusion}
\label{sec:sec4}

In this paper, we proposed QAEMCMC, where QA was utilized to generate proposal states in the MCMC subroutine.
We investigated the performance of QAEMCMC compared to the classical MCMC in the SK model.
We exhibited that QAEMCMC yielded a larger spectral gap than the classical updates.
The QA update mitigated the difficulty of sampling in the low-temperature regime.
More than doubled quantum enhancement existed from the exponents of the spectral gap. 
We also showed the dynamical simulation of the averaged energy and the total variation distance between the empirical distribution and target distribution.
We obtained a better estimation of observables and the distribution statistics faster than the classical MCMC by QAEMCMC.
The sample characteristics obtained by QAEMCMC demonstrated that QA could explore broader regions of the configuration with low energy and overcome trapping in local minima.
Our results exhibited that the proposal distribution obtained by QA was effective in the sampling tasks for complex systems. 
They may be useful to verify the validity of the usage of QA data in NMCMC \cite{Giuseppe_2023}.

It is noted that QA cannot find the degenerated ground states with equal probability \cite{Matsuda_2009, Konz2019}.
The probability of the isolated ground state is suppressed in the large annealing time region due to the bias generated from the transverse field term.
The higher-order driver term mitigates this unfair sampling of QA. In the intermediate annealing time region, the probability of the isolated ground state remains a finite value. If we choose the proper annealing time, the unfair sampling problem can be relieved. This unfair sampling problem does not preclude QA applications from being included in the MCMC subroutine. 
In a recent study \cite{christmann_2024}, time-dependent Hamiltonian evolution like \textit{reverse annealing} \cite{yamashiro_pra_2019} has been applied to QEMCMC.
Our approach is similar to this approach. 
The reverse annealing leads to exploring the region around the initial spin configuration. 
The QA update can yield a global update irrespective of the current spin configuration, which creates a large spectral gap in the low-temperature region. A detailed study is needed since the large gap between the two approaches exists for the spectral gap in low-temperature regions. Our simulation was limited to the small system size, and whether QAEMCMC is effective for large-size problems is an open question. As conducted in the previous study \cite{christmann_2024}, 
tensor networks are useful for validating the efficacy of QAEMCMC for large-scale problems.
This is an interesting future research topic.
\bibliography{main.bib}
\section*{Acknowledgements}
The authors are grateful to Hidetoshi Nishimori for many fruitful discussions.
This paper was partly based on results obtained from a project, JPNP16007, commissioned by the New Energy and Industrial Technology Development Organization (NEDO), Japan.
This work was partly performed for Council for Science, Technology and Innovation (CSTI), Cross-ministerial Strategic Innovation Promotion Program (SIP), "Promoting the application of advanced quantum technology platforms to social issues"(Funding agency : QST).
\section*{Author contributions statement}
S.A. designed this study and conducted the numerical experiments.
T.K. contributed to managing this project.
All authors discussed the results and commented on the manuscript. 
\section*{Competing Interests}
The author(s) declare no competing interests.
\section*{Data Availability}
 The datasets used in our study are available from the corresponding author upon reasonable
request.
\end{document}